\documentclass[final]{cvpr}

\usepackage{times}
\usepackage{epsfig}
\usepackage{graphicx}
\usepackage{amsmath}
\usepackage{amssymb}
\usepackage{booktabs, multirow} 
\usepackage{soul}
\usepackage[table]{xcolor} 
\usepackage{makecell}
\usepackage{float}

\usepackage[pagebackref=true,breaklinks=true,colorlinks,bookmarks=false]{hyperref}

\def\cvprPaperID{****} 
\def\confYear{CVPR 2021}

\begin{document}

\title{Soft-Attention Improves Skin Cancer Classification Performance}

\author{Soumyya Kanti Datta \quad Seyed Mohammad Abuzar Hashemi \quad Sargur N Srihari \quad Mingchen Gao \\
State University of New York, Buffalo\\
Buffalo, New York, USA\\
{\tt\small \{soumyyak,srihari,mgao8\}@buffalo.edu \quad ma.hashemi.786@gmail.com}}




\maketitle

\begin{abstract}

In clinical applications, neural networks must focus on and highlight the most important parts of an input image. Soft-Attention mechanism enables a neural network to achieve this goal. This paper investigates the effectiveness of Soft-Attention in deep neural architectures. The central aim of Soft-Attention is to boost the value of important features and suppress the noise-inducing features. We compare the performance of VGG, ResNet, Inception ResNet v2 and DenseNet architectures with and without the Soft-Attention mechanism, while classifying skin lesions. The original network when coupled with Soft-Attention outperforms the baseline\cite{rezvantalab2018dermatologist} by 4.7\% while achieving a precision of 93.7\% on HAM10000 dataset \cite{ham10000}. Additionally, Soft-Attention coupling improves the sensitivity score by 3.8\% compared to baseline\cite{zhang2019attention} and achieves  91.6\% on ISIC-2017 dataset \cite{ISIC-2017}. The code is publicly available at github\footnote{\url{https://github.com/skrantidatta/Attention-based-Skin-Cancer-Classification}}.
\end{abstract}

\section{Introduction}

Skin cancer is the most common cancer and one of the leading causes of death worldwide.
Every day, more than 9500 people\footnote{\url{https://www.skincancer.org/skin-cancer-information/skin-cancer-facts/}} in the United States are diagnosed with skin cancer, with 3.6 million people\footnote{\url{https://www.skincancer.org/skin-cancer-information/basal-cell-carcinoma/}} diagnosed with basal cell skin cancer each year.
Early diagnosis of the illness has a significant effect on the patients' survival rates.
As a result, detecting and classifying skin cancer is important. 

It is difficult to distinguish between malignant and benign skin diseases because they look so similar. Although a dermatologist's visual examination is the first step in detecting and diagnosing a suspicious skin lesion, it is usually followed by dermoscopy imaging for further analysis \cite{zunair2020melanoma}. Dermoscopy images provide a high-resolution magnified image of the infected skin region, but they are not without their drawbacks. Due to the image size being large, it becomes difficult for the feature extractors to extract out the relevant features for classification. Various methods such as Segmentation and detection, Transfer learning, General Adversarial networks, etc. have been used to detect and classify skin cancer. Despite significant progress, skin cancer classification is still a difficult task.
This is due to the lack of annotated data and low inter-class variation. Furthermore, the task is complicated by contrast variations, color, shape, and size of the skin lesion, as well as the presence of various artifacts such as hair and veins.  Inspired by the work done in \cite{Shaikh_2020}, this paper studies the effect of soft attention mechanism in deep neural networks. Deep learning architectures identify the image class by learning the salient features and nonlinear interactions.
The soft-attention mechanism improves performance by focusing primarily on relevant areas of the input.
Moreover, the soft-attention mechanism makes the image classification process transparent to medical personnel, as it maps the parts of the input that the network uses to classify the image, thereby, increasing trust in the classification model.
\begin{figure}[t]
\centering
\includegraphics[width=1.0\linewidth]{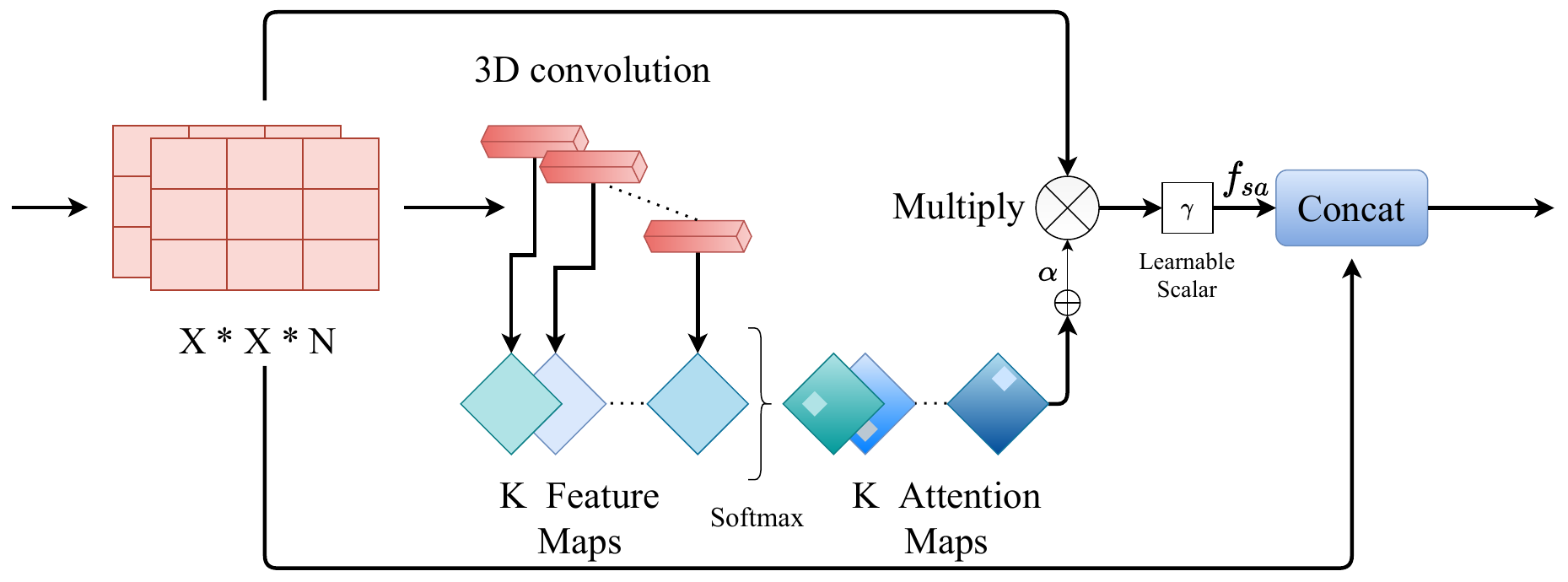}
\caption{Soft Attention unit}
\label{fig:attention}
\end{figure}
\section{Related Work}

Following Krichevsky\cite{krizhevsky2017imagenet}, large-scale image classification tasks using deep convolutional neural networks have become common.
As reported in the paper\cite{esteva2017dermatologist}, the task of skin cancer classification using images has improved rapidly since the implementation of Deep Neural Networks.
To make progress, we suggest that soft attention be used to identify fine-grained variability in the visual features of skin lesions. 

Existing art in the field of skin cancer classification used streamlined pipelines based upon current Computer Vision. \cite{fornaciali2016towards}. Masood et al. in their paper.\cite{masood2013computer} proposed a general framework from the viewpoint of computer vision, where the methods such as calibration, preprocessing, segmentation, balancing of classes and cross validation are used for automated melanoma screening. In 2018, Valle et al.\cite{valle2020data} investigated ten different methodologies to evaluate deep learning models for skin lesion classification. 
Data augmentation, model architecture, image resolution, input normalization, train dataset, use of segmentation, test data augmentation, additional use of support vector machines, and use of transfer learning are among the ten methodologies they evaluated.
They stated that data augmentation had the greatest impact on model efficiency.
The same observation is confirmed by Perez's 2018 paper "Data Augmentation for Skin Lesion Analysis"\cite{perez2018data}.

Nonetheless, the problems of low inter-class variance and class imbalance in skin lesion image datasets remain, seriously limiting the capabilities of deep learning models\cite{yu2016automated}.
To fix the lack of annotated data, Zunair et al.\cite{zunair2020melanoma} proposed the use of adversarial training and Bissoto et al.\cite{bissoto2018skin} proposed the use of Generative Adversarial Networks to produce realistic synthetic skin lesion photos.

\section{Experiment Settings And Method}

In this paper, five deep neural networks which are ResNet34, ResNet50 \cite{he2016deep}, Inception ResNet v2\cite{szegedy2016inception}, DenseNet201\cite{huang2017densely}  and VGG16 \cite{simonyan2014very}, are implemented with soft attention mechanism, to classify skin cancer images. ResNet34, ResNet50\cite{he2016deep}, Inception ResNet v2,  DenseNet201\cite{huang2017densely} and VGG16\cite{simonyan2014very} are all state of the art feature extractors which are trained on ImageNet dataset. The main components and architecture of the proposed approach is described below:

\subsection{Dataset}
The experiment is performed on two datasets separately. The two datasets are as follows: HAM10000 dataset \cite{ham10000} and ISIC 2017 dataset.

The HAM10000 dataset \cite{ham10000} consists of 10015 dermatoscopic images of a size of 450 × 600. It consists of 7 diagnostic categories as follows: Melanoma(MEL), Melanocytic Nevi(NV), Basal Cell Carcinoma(BCC), Actinic Keratosis, and Intra-Epithelial Carcinoma(AKIEC), Benign Keratosis(BKL), Dermatofibroma(DF), Vascular lesions(VASC). All the images are resized to 299 x 299 for Inception ResNet v2\cite{szegedy2016inception} architecture and 224 x 224 for the other architectures.
\begin{figure}[h]
\centering
\includegraphics[width=1.0\linewidth]{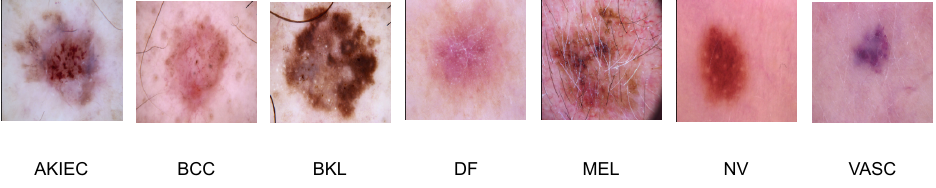}
\caption{Example of Skin lesions in HAM10000 dataset \cite{ham10000}}
\end{figure}

The ISIC 2017 dataset consists 2600 images of size 767 x 1022.
In the training dataset there are 2000 images of 3 catagories as follows: benign nevi, seborrheic keratosis, and melanoma. The test dataset consist of 600 images. In this experiment we are training our model to classify only benign nevi and seborrheic keratosis. All the images resized to 224 x 224.

The data in both datasets is then cleaned to remove class imbalances. This is done by the process of over-sampling and under-sampling of data so that there are equal number of images per class. The images are then normalized by dividing each pixel with 255 to keep the pixel values in the range 0 to 1.

\subsection{Soft Attention}
When it comes to skin lesion images, only a small percentage of pixels are relevant as the rest of the image is filled with various irrelevant artifacts such as veins and hair. So, to focus more on these relevant features of the image, soft attention is implemented. Inspired by the work proposed by Xu et al \cite{xu2015show}, for image caption generation and the work done by Shaikh et al \cite{Shaikh_2020}, where they used attention mechanism on images for handwriting verification, in this paper, soft attention is used to classify skin cancer. 
\begin{figure}[h]
\centering
\includegraphics[width=1.0\linewidth]{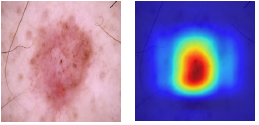}
\caption{Images with Soft Attention}
\label{fig:2}
\end{figure}

In Figure [\ref{fig:2}], we can see that areas with higher attention are red in color . This is because soft attention discredits irrelevant areas of the image by multiplying the corresponding feature maps with low weights. Thus the low attention areas have weights closer to 0. With more focused information, the model performs better.

In the soft attention module as discussed in paper \cite{Shaikh_2020} and \cite{tomita2019attention}, the feature tensor (t) which flows down the deep neural network is used as input.
\begin{equation}
     f_{sa} = \gamma t((\sum_{k=1}^{K}softmax(W_{k}*t)))
\end{equation}

This feature tensor $t \in \mathbb{R}^{h \times w \times d}$ is input to a 3D convolution layer\cite{tran2015learning} with weights $W_k \in \mathbb{R}^{h\times w\times d \times K}$, where $K$ is the number of 3D weights. The output of this convolution is normalized using softmax function to generate $K = 16$ attention maps. As shown in Figure \ref{fig:attention}, these attention maps are aggregated to produce a unified attention map that acts as a weighting function $\alpha$. This $\alpha$ is then multiplied with $t$ to attentively scale the salient feature values, which is further scaled by $\gamma$ a learnable scalar. Finally, the attentively scaled features ($f_{sa}$) are concatenated with the original feature $t$ in form of a residual branch. During training we initialize $\gamma$ from $0.01$ so that the network can slowly learn to regulate the amount of attention required by the network.
 
\subsection{Model Setup}
In this section, the detailed architecture of all the models is discussed.
For all experiments, to train the networks, Adam optimizer\cite{kingma2014adam} of 0.01 learning rate and 0.1 epsilon is used. A batch normalization\cite{ioffe2015batch} layer is added after each layer in all the networks to introduce some regularization. For the HAM10000 dataset \cite{ham10000}, since there are 7 classes of skin cancer, an output layer with 7 hidden units is implemented, followed by a softmax activation unit. All the experiments were executed on the Keras framework.  

 \begin{figure}[h]
\centering
\includegraphics[width=1.0\linewidth]{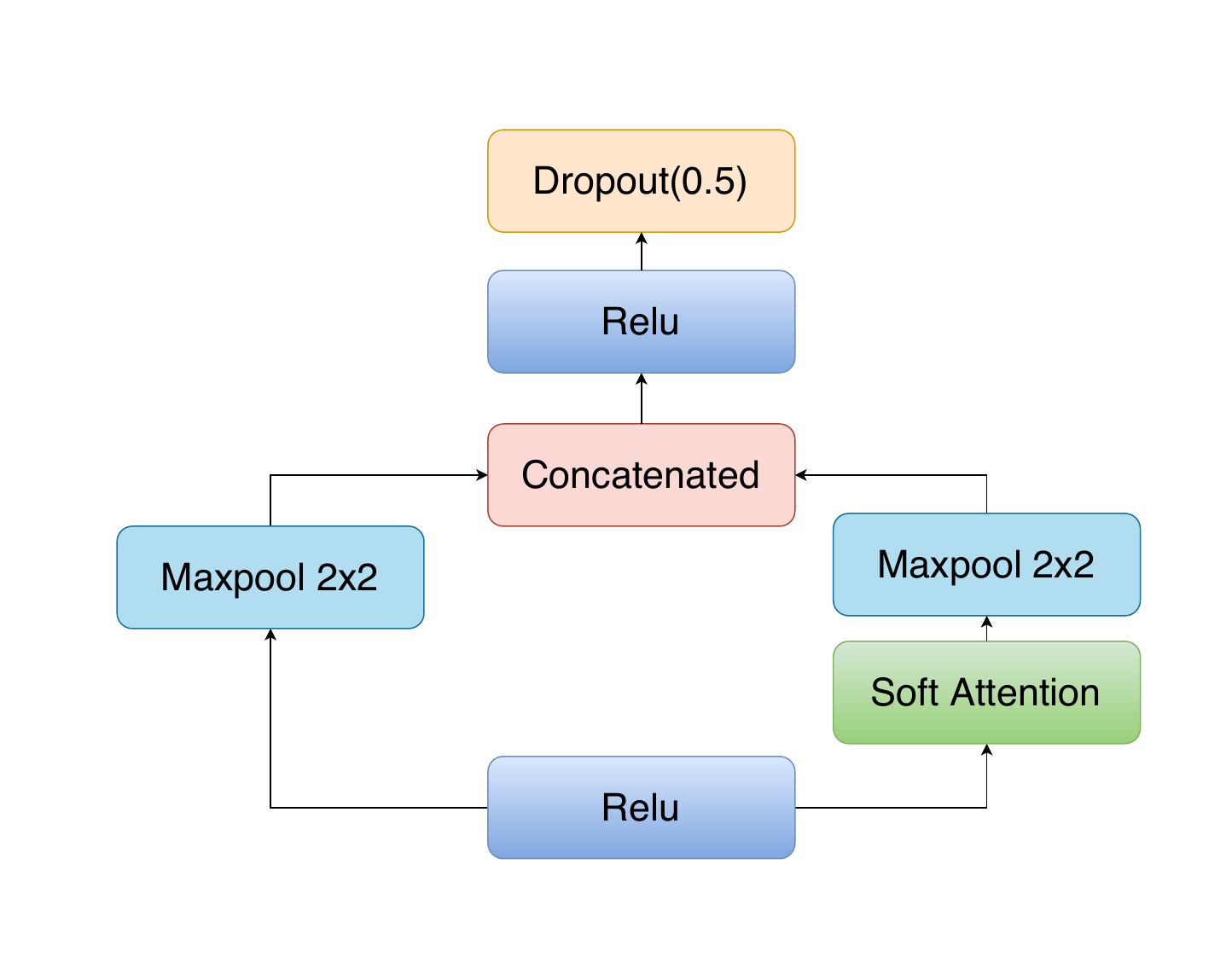}
\caption{The schema for Soft Attention Block }
\label{fig:4}
\end{figure} 

  \subsubsection{Inception ResNet v2}
\begin{figure}[h]
\centering
\includegraphics[height=1.0\linewidth]{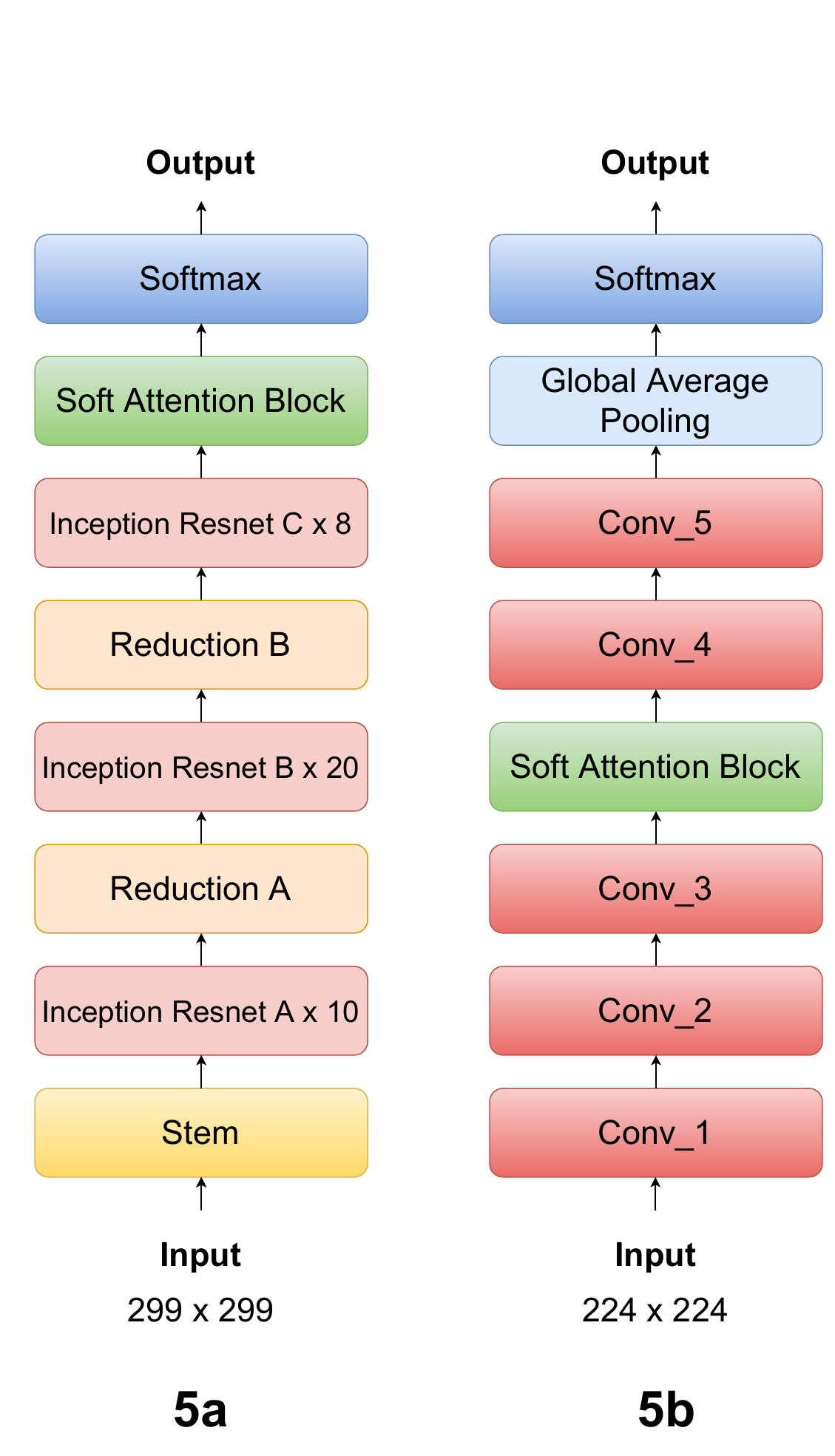}
\caption{\textbf{5a}. End to end architecture Of Inception ResNet v2\cite{szegedy2016inception} with Soft Attention Block. \textbf{5b}. End to end architecture Of  ResNet34\cite{he2016deep} with Soft Attention Block . conv\textunderscore x indicates convolution blocks, where x is the block number.}
\label{fig:5}
\end{figure}

  In Inception ResNet v2\cite{szegedy2016inception}, the soft attention layer is added to the Inception Resnet C block of the model where the feature size of the image is 8 x 8  as shown in Figure [\ref{fig:5}a]. In this case, the soft attention layer is followed by a maxpool layer with a pool size of 2x2, which is then concatenated with the filter concatenate layer of the inception block. The concatenate layer is then followed by a relu activation unit. To regularize the output of the attention layer, the activation unit is followed by a 0.5 dropout layer\cite{srivastava2014dropout} as in Figure [\ref{fig:4}]. The network is trained for 150 epochs with early stopping patience of 30. The overall network is shown in Figure [\ref{fig:5}a].
 
\subsubsection{DenseNet201}
\begin{figure}[h]
\centering
\vspace{5mm}
\includegraphics[height=\linewidth,angle=270]{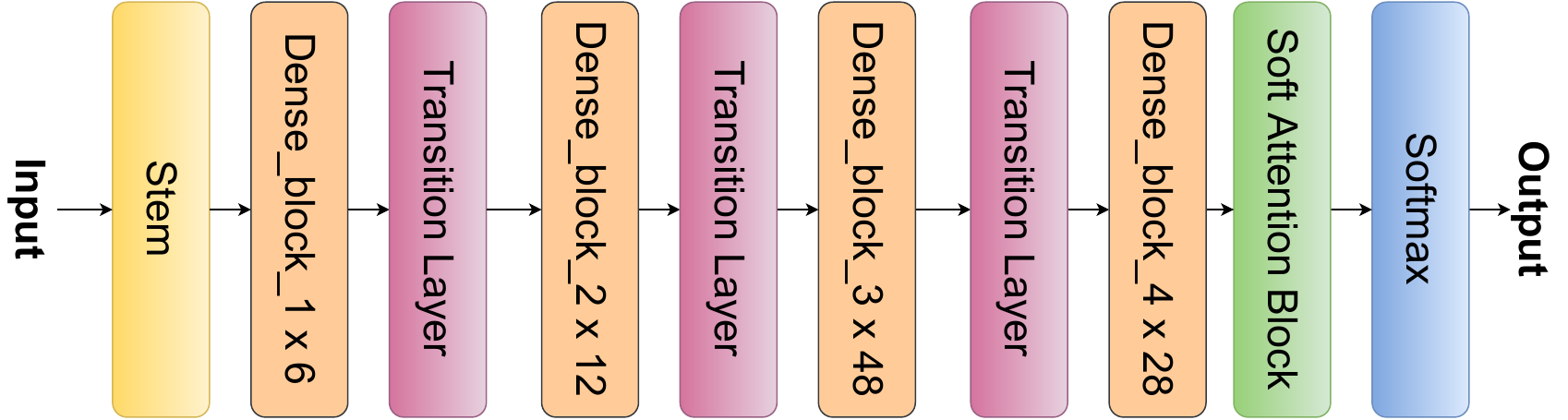}
\vspace{5mm}
\caption{End to end schema of DenseNet201\cite{huang2017densely} with Soft Attention Block.}
\label{fig:6}
\end{figure}

  In DenseNet201\cite{huang2017densely}, the soft attention layer is added to the 4th dense block where the size of feature map of the image is 7 x 7  as shown in Figure[\ref{fig:6}]. Like in the previous model, the soft attention layer is integrated with the same procedure as it was integrated with the  Inception ResNet V2\cite{szegedy2016inception} architecture.[\ref{fig:4}]. The network is trained for 150 epochs with early stopping patience of 35.

\subsubsection{ResNet34 and ResNet50}
 \begin{figure*}[!htp]
\centering
\includegraphics[width=0.90\linewidth]{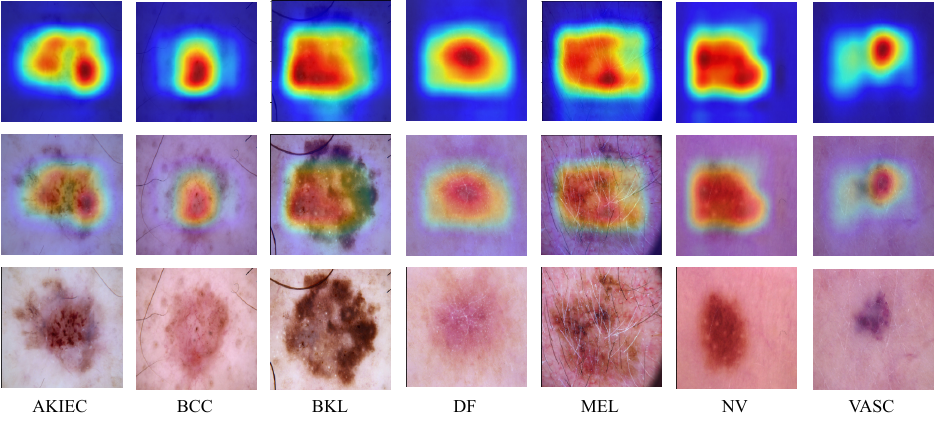}
\caption{Soft Attention maps of Skin lesion in Inception ResNet V2 on HAM10000 dataset \cite{ham10000}}
\label{fig:attention_map}
\end{figure*}

 In ResNet34\cite{he2016deep}, a soft attention layer is added after the 3rd convolution block where the size of feature map is 28 x 28 as shown in [\ref{fig:5}b] whereas, in the ResNet50\cite{he2016deep}, the soft attention layer is added after the 5th convolution block where the size of feature map is 7 x 7. In both cases, the soft attention layer is followed by a maxpool layer with a pool size of 2x2, which is then concatenated with the standard maxpool layer of the architecture, as shown in Figure [\ref{fig:4}]. The concatenate layer is then followed by a relu activation unit. To regularize the output of the attention layer, the activation unit is followed by a 0.5 dropout\cite{srivastava2014dropout} layer. This is the same approach as to how the soft attention module was integrated with the Inception ResNet V2\cite{szegedy2016inception} architecture. The overall architecture for ResNet 34 model is shown in Figure [\ref{fig:5}b].

 \subsubsection{VGG16}
\begin{figure}[h]
\centering
\vspace{5mm}
\includegraphics[height=\linewidth,angle=270]{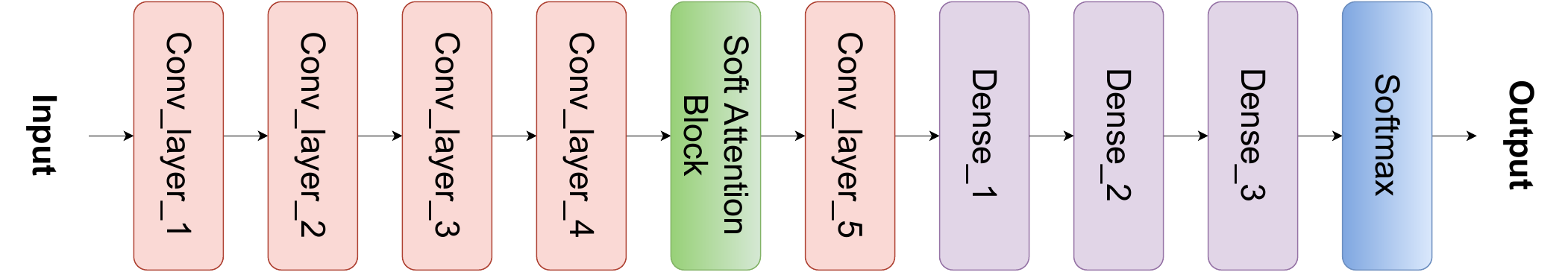}
\vspace{5mm}
\caption{End to end schema of VGG16\cite{simonyan2014very} with Soft Attention Block. conv x indicates convolution layer with x filters.}
\label{fig:7}
\end{figure}

  In VGG16\cite{simonyan2014very}, the soft attention layer is added after the conv\_layer\_4 of the VGG16 architecture where the size of feature map is 28 x 28. Like in the previous model, the soft attention layer is integrated with the same procedure as it was integrated with the ResNet\cite{he2016deep} and Inception ResNet V2\cite{szegedy2016inception} architecture.[\ref{fig:4}]. The network is trained for 300 epochs with early stopping patience of 65. The overall architecture for the model is shown in Figure [\ref{fig:7}].

In Figure [\ref{fig:7}], a Conv\_layer block consists of two to three convolution layers with filters of sizes ranging from 64 to 512, followed by a maxpool layer. Conv\_layer\_1, and Conv\_layer\_2 consists of two convolution layers each with 64, and 128 filters respectively, and  Conv\_layer\_3, Conv\_layer\_4 and  Conv\_layer\_5 consists of three convolution layers each with 256, 512 and 512 filters respectively.

\subsection{Loss Function}
In this experiment, there are seven different classes of skin cancer. Hence , categorical cross entropy loss \((L_{CCE})\) is used to optimize the  neural network.
\begin{align}
    L_{CCE} = -\sum_{i=1}^C t_ilog(f(s)_i)
\end{align}
where
\begin{align}
   f(s)_i = \frac{e^{s_i}}{\sum_{j=1}^C e^{s_j}}
\end{align}
Here, as there are seven classes, \(C \in [0..6]\), where \(t_i\) is the ground truth and \(s_i\) is the CNN score for each class i in C. \({f(s)_i}\) is the softmax activation function applied to the scores.

\begin{table*}[!htp]\centering
\resizebox{\textwidth}{!}{
    \begin{tabular}{l|cc|cc|cc|cc|cc|cc|cc|cc|cc|cc|cc}\toprule
    \multirow{2}{*}{\textbf{Dis.}} &\multicolumn{10}{c}{\textbf{Precision}} &\multicolumn{10}{|c|}{\textbf{AUC}} &\multirow{2}{*}{\textbf{\makecell{\#}}} \\\cmidrule{2-21}
    &\textbf{\cite{szegedy2016inception}} &\textbf{\makecell{
\cite{szegedy2016inception} +\\SA
}} &\textbf{\cite{huang2017densely}} &\textbf{\makecell{\cite{huang2017densely} +\\SA
}} &\textbf{\cite{simonyan2014very}} &\textbf{\makecell{\cite{simonyan2014very} +\\SA
}} &\textbf{\cite{he2016deep}50} &\textbf{\makecell{\cite{he2016deep}50 +\\SA
}} &\textbf{\cite{he2016deep}34} &\textbf{\makecell{
\cite{he2016deep}34 +\\SA
}} &\textbf{\cite{szegedy2016inception}} &\textbf{\makecell{\cite{szegedy2016inception} +\\SA}} &\textbf{\cite{huang2017densely}} &\textbf{\makecell{\cite{huang2017densely} +\\SA}} &\textbf{\cite{simonyan2014very}} &\textbf{\makecell{\cite{simonyan2014very} +\\SA}} &\textbf{\cite{he2016deep}50} &\textbf{\makecell{\cite{he2016deep}50 +\\SA
}} &\textbf{\cite{he2016deep}34} &\textbf{\makecell{\cite{he2016deep}34 +\\SA
}} & \\\midrule
    AKIEC &0.830 &\textbf{1.000} &\textbf{1.000} &0.920 &0.620 &0.700 &0.740 &0.670 &0.670 &0.500 &\textbf{0.993} &0.981 &0.975 &0.967 &0.949 &0.964 &0.980 &0.981 &0.969 &0.970 &23 \\
    BCC &0.850 &0.880 &0.830 &0.800 &0.540 &0.620 &\textbf{0.910} &0.880 &0.660 &0.880 &0.997 &\textbf{0.998} &0.993 &0.994 &0.977 &0.984 &0.997 &0.996 &0.991 &0.993 &26 \\
    BKL &\textbf{0.850} &0.720 &0.690 &0.730 &0.570 &0.630 &0.670 &0.670 &0.510 &0.520 &0.970 &\textbf{0.982} &0.960 &0.964 &0.930 &0.900 &0.948 &0.964 &0.904 &0.916 &66 \\
    DF &0.670 &\textbf{1.000} &0.500 &\textbf{1.000} &0.250 &0.500 &0.800 &\textbf{1.000} &0.400 &0.330 &0.973 &\textbf{0.982} &0.851 &0.921 &0.847 &0.809 &0.973 &0.971 &0.925 &0.949 &6 \\
    MEL &0.700 &0.670 &0.540 &0.530 &0.500 &0.430 &0.520 &\textbf{0.730} &0.420 &0.540 &0.965 &0.974 &0.963 &\textbf{0.976} &0.925 &0.956 &0.961 &0.973 &0.910 &0.953 &34 \\
    NV &0.930 &\textbf{0.970} &0.950 &0.950 &0.930 &0.950 &0.950 &0.950 &0.930 &0.930 &\textbf{0.984} &\textbf{0.984} &0.975 &0.976 &0.954 &0.951 &0.974 &0.979 &0.944 &0.958 &663 \\
    VASC &\textbf{1.000} &\textbf{1.000} &0.900 &0.830 &\textbf{1.000} &\textbf{1.000} &0.900 &\textbf{1.000} &0.910 &0.820 &\textbf{1.000} &\textbf{1.000} &0.993 &0.999 &0.972 &0.999 &0.995 &0.999 &0.999 &0.996 &10 \\
    \midrule
    Avg &0.832 &\textbf{0.892} &0.771 &0.824 &0.631 &0.690 &0.783 &0.841 &0.642 &0.646 &0.983 &\textbf{0.984} &0.959 &0.971 &0.936 &0.937 &0.975 &0.980 &0.949 &0.962 &828 \\\midrule
    W. Avg &0.905 &\textbf{0.937} &0.904 &0.909 &0.862 &0.882 &0.898 &0.910 &0.857 &0.865 &0.982 &\textbf{0.984} &0.974 &0.975 &0.951 &0.948 &0.972 &0.978 &0.942 &0.957 &828 \\
    \bottomrule
    \end{tabular}
    
}
\vspace{1mm}
\caption{Ablation results for choosing the best model on HAM10000 dataset \cite{ham10000}. {\cite{szegedy2016inception}} refers to IRv2 architecture,   {\cite{huang2017densely}} refers to DenseNet 201 architecture,  
{\cite{simonyan2014very}} refers to VGG 16 architecture, and 
{\cite{he2016deep}} refers to ResNet architecture.
}
\label{tab:ablation}

\end{table*}

\subsection{Evaluation Metrics}
In this paper, the model is evaluated using  $Precision = \frac{TP}{TP+FP}$, $Accuracy = \frac{TP+TN}{T}$, $Sensitivity = \frac{TP}{TP+FN}$, $Specificity = \frac{TN}{TN+FP}$ and AUC scores\cite{AUC}. Here TN, TP, FP, FN, T mean, True Negatives, True Positives, False Positives, False Negatives, Total Number respectively.
\section{Discussion}
\subsection{Ablation Analysis}
Table \ref{tab:ablation} lists, the performance of all the models in terms of precision, and AUC score on HAM10000 dataset \cite{ham10000}. In this table (+SA) stands for models with soft attention. IRv2 stands for Inception ResNet v2\cite{szegedy2016inception}, \cite{he2016deep}34 stands for ResNet34\cite{he2016deep} and \cite{he2016deep}50 stands for ResNet50\cite{he2016deep}. From the table, it can be observed that IRv2 when coupled with SA (IRv2+SA) shows significant improvements in results, with a precision and AUC score of 93.7\% and 98.4\% respectively, which are also the highest scores amongst all models. Furthermore, we can see that Soft Attention (SA) boosts the performance of IRv2 by 3.2\% in terms of precision as compared to the original IRv2 model. This phenomenon is true for VGG16, ResNet34, ResNet50 and DenseNet201 as well. For instance, Soft Attention (SA) boosts the precision of DenseNet201\cite{huang2017densely}, ResNet34\cite{he2016deep}, ResNet50\cite{he2016deep},  and VGG16\cite{simonyan2014very} by 0.5\%, 0.8\%, 1.2\% and 2\% respectively. We see a similar  behaviour for the AUC scores when SA block is integrated in to the networks, such as, the performance of  ResNet50\cite{he2016deep}, and ResNet34\cite{he2016deep} has grown by 0.6\% and 1.5\% respectively and the performance of  DenseNet201\cite{huang2017densely}, and  VGG16\cite{simonyan2014very} is on par with the original models.

Although IRv2+SA performs the best in terms of weighted average(W.Avg), when we look at it's class wise performance, we can see that Soft Attention enhances the efficiency of the original IRv2 while categorizing AKIEC, BCC, DF and NV by 17\%, 3\%, 33\% and 4\% respectively in terms of precision. Moreover, when comparing AUC scores, the IRv2+SA performs better for BKL and MEL by 1.2\% and 0.9\% respectively, while, for BCC, NV and VASC, IRv2+SA performs as good as original model. 

We thus select IRv2 coupled with SA (IRv2+SA) for our experiments, also the SA block consistently boosts the performance of it's original counterpart, hence, we can justify the integration of Soft Attention to the networks.

\begin{figure*}[t]
\centering
\includegraphics[width=1.0\linewidth]{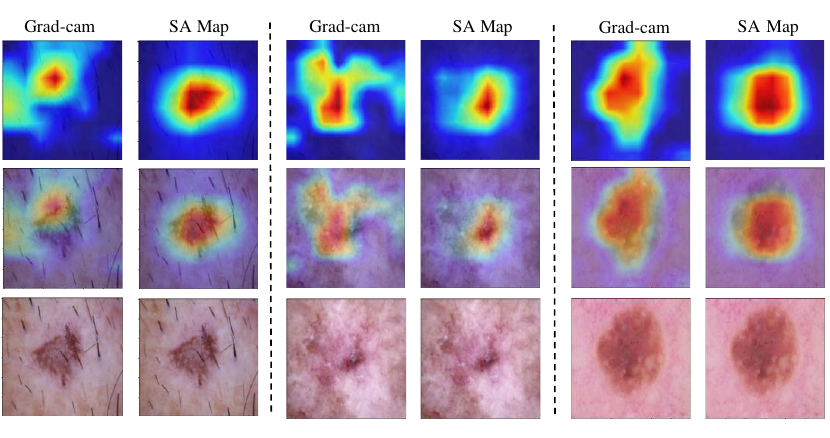}
\vspace{-8mm}
\caption{Comparison of GradCAM \cite{Grad-cam} heatmaps with our Soft Attention (SA) maps on HAM10000 dataset \cite{ham10000}}
\label{fig:compare}
\end{figure*}


\subsection{Quantitative Analysis}
When we tested the model with different train-test splits on the HAM10000 dataset \cite{ham10000}, we discovered that the model with 85 \% training data outperforms the model with 80 \% and 70 \% training data by 2.2 \% and 2.6 \% respectively, as shown in Table \ref{tab:split}. Hence we select 85/15\% training/testing split for performing our experiments.

\begin{table}[H]\centering

\resizebox{\columnwidth}{!}{
\begin{tabular}{lcccccccccc}\toprule
&\multicolumn{3}{c}{\textbf{split = 15}} &\multicolumn{3}{c}{\textbf{split = 20}} &\multicolumn{3}{c}{\textbf{split = 30}} \\\cmidrule{2-10}
Disease &Support &Precision &AUC &Support &Precision &Auc &Support &Precision &Auc \\\midrule
AKIEC &23 &1.000 &0.981 &30 &0.750 &0.958 &45 &0.690 &0.972 \\
BCC &26 &0.880 &0.998 &35 &0.880 &0.992 &53 &0.830 &0.995 \\
BKL &66 &0.720 &0.982 &88 &0.790 &0.972 &132 &0.720 &0.960 \\
DF &6 &1.000 &0.973 &8 &1.000 &0.998 &12 &0.710 &0.989 \\
MEL &34 &0.670 &0.974 &46 &0.490 &0.953 &69 &0.600 &0.946 \\
NV &663 &0.970 &0.984 &883 &0.960 &0.981 &1325 &0.960 &0.978 \\
VASC &10 &1.000 &1.000 &13 &0.920 &0.999 &19 &0.860 &0.983 \\
\midrule
Avg &828 &0.892 &0.984 &1103 &0.827 &0.9793 &1655 &0.766 &0.975 \\
\midrule
W. Avg &828 &\textbf{0.937} &0.984 &1103 &0.915 &0.9797 &1655 &0.911 &0.976 \\
\bottomrule
\end{tabular}
}
\vspace{1mm}
\caption{Comparison with Models with different train-test split on HAM10000 dataset \cite{ham10000}}\label{tab:split}
\end{table}

Furthermore, the proposed approach is compared with state-of-the-art models for skin cancer classification on the HAM10000 dataset \cite{ham10000} in Table \ref{tab:compare}. Our Soft Attention-based approach outperforms the baseline\cite{rezvantalab2018dermatologist} by 4.7\% in terms of precision. In terms of AUC scores, our Soft Attention-based approach clearly outperforms them all by 0.5\% to 4.3\%. 
\begin{table}[H]\centering

\resizebox{\columnwidth}{!}{
\begin{tabular}{lccc}\toprule
\textbf{Model} &\textbf{Avg AUC}& \textbf{Precision} & \textbf{Accuracy}\\\midrule

Loss balancing and ensemble\cite{gessert2020skin} &0.941 &-&0.926\\
Single Model Deep Learning\cite{yao2021single}  &0.974&- & 0.864\\
Data classification augmentation\cite{shen2021low}  &0.975&- &0.853\\
Two path CNN model\cite{nadipineni2020method} &-&-& 0.886 \\
Various Deep CNN (Baseline) \cite{rezvantalab2018dermatologist} &0.979 &0.890&- \\
\midrule
\textbf{IRv2+SA(Proposed Approach) } &\textbf{0.984} &\textbf{0.937} &\textbf{0.934} \\

\bottomrule
\end{tabular}
}
\vspace{1mm}
\caption{Comparison with state-of-the-art-Model in terms of Average AUC score on HAM10000 dataset \cite{ham10000}}\label{tab:compare}

\end{table}

In Table \ref{tab:ISIC2017}, the performance of the proposed approach Inception Resnet V2\cite{szegedy2016inception} (IRv2\textsubscript{5x5}+SA and IRv2\textsubscript{12x12}+SA)  with soft attention is measured on ISIC-2017 dataset \cite{ISIC-2017} on basis of AUC scores, Accuracy , Sensitivity and Specificity   with the state-of-the-art models. 
\begin{table}[H]\centering
\resizebox{\columnwidth}{!}{
\begin{tabular}{lccccc}\toprule
\textbf{Networks} &\textbf{AUC} &\textbf{Accuracy} &\textbf{Sensitivity} &\textbf{Specificity} \\\midrule
ResNet50 \cite{he2016deep} &0.948 &0.842 &0.867 &0.837 \\
RAN50 \cite{wang2017residual}&0.942 &0.862 &0.878 &0.859 \\
SEnet50 \cite{hu2018squeeze}&0.952 &0.863 &0.856 &0.865 \\
ARL-CNN50\cite{zhang2019attention} &0.958 &0.868 &0.878 &\textbf{0.867} \\
IRv2\textsubscript{12x12}+SA &0.935 &0.898 &\textbf{0.945} &0.711 \\
\midrule
\textbf{IRv2\textsubscript{5x5}+SA } &\textbf{0.959} &\textbf{0.904} &0.916 &0.833 \\
\bottomrule
\end{tabular}
}
\vspace{1mm}
\caption{Comparison with state-of-the-art-Model in terms of AUC, Accuracy, sensitivity and specificity score on ISIC-2017 dataset \cite{ISIC-2017}}\label{tab:ISIC2017}
\end{table}

From Table \ref{tab:ISIC2017}, it can be observed that in IRv2\textsubscript{5x5}+SA, and  in IRv2\textsubscript{12x12}+SA, the attention layer was added when the feature map size is 5x5 and 12x12 respectively.
Out of the two models with soft attention, the model IRv2\textsubscript{5x5}+SA outperforms IRv2\textsubscript{12x12}+SA in terms of AUC scores, Accuracy, and Specificity by a percentage of 2.4\%, 0.6\%, and 12.2\% respectively whereas IRv2\textsubscript{12x12}+SA outperforms IRv2\textsubscript{5x5}+SA in terms of Sensitivity by 2.9\%. In this case, the attention layer was added when the feature size is 5x5. When IRv2\textsubscript{5x5}+SA  is compared with the ARL-CNN50\cite{zhang2019attention} (baseline model), it performs on par with it in terms of AUC score but our model outperforms it when it comes to accuracy and Sensitivity by 3.6\% and 3.8\% respectively. But ARL-CNN50\cite{zhang2019attention} takes the upper hand when it comes to Specificity by 3.4\%. Since sensitivity measures the proportion of correctly identified positives and specificity measures the proportion of correctly identified negatives, we are prioritizing Sensitivity because classifying a person with cancer as not having cancer is riskier than vice versa.

\subsection{Qualitative Analysis}

Fig.\ref{fig:attention_map} displays the Soft Attention heat maps from the IRv2+SA model.  In the Fig.\ref{fig:attention_map}, the images on the bottom row are the input images of the seven skin cancer categories. The images in the middle row show the Soft Attention maps superimposed on input images to show where the model is focusing and the images of the top row are attention maps themselves.

In Fig.\ref{fig:compare}, we show pairs of comparison between the Soft Attention maps with Grad-CAM \cite{Grad-cam} heatmaps. 
In the first pair, the SA map focuses on the main part of the lesion area whereas the Grad-cam heatmap is slightly shifted towards top left and is also spread out on the uninfected area of skin. We have similar observations for the second and third pairs as well. From this observation it is evident that the Soft Attention maps are focused more on the relevant locations of the image compared to Grad-CAM\cite{Grad-cam} heatmaps. 
\section{Conclusion}

In this paper, we present the implementation and utility of Soft Attention mechanism being applied while image encoding to tackle the problem of high-resolution skin cancer image classification.
The model outperformed the current state-of-the-art approaches on the HAM10000 dataset \cite{ham10000} and the ISIC-2017 dataset \cite{ISIC-2017}.
This demonstrates the Soft Attention based deep learning architecture's potential and effectiveness in image analysis. 
The Soft Attention mechanism also eliminates the need of using external mechanisms like GradCAM \cite{Grad-cam}, and internally provides the location of where the model focuses while categorizing a disease, while also boosting the performance of the main network.  Soft Attention has the added advantage of naturally dealing with image noise internally.
In the future, this model can be implemented in dermoscopy systems to assist dermatologists. This mechanism can be easily implemented to classify data from other medical databases as well.

{\small
\bibliographystyle{ieee_fullname}
\bibliography{egbib}

@article{Shaikh_2020,
   title={Attention based Writer Independent Verification},
   ISBN={9781728199665},
   url={http://dx.doi.org/10.1109/ICFHR2020.2020.00074},
   DOI={10.1109/icfhr2020.2020.00074},
   journal={2020 17th International Conference on Frontiers in Handwriting Recognition (ICFHR)},
   publisher={IEEE},
   author={Shaikh, Mohammad Abuzar and Duan, Tiehang and Chauhan, Mihir and Srihari, Sargur N.},
   year={2020},
   month={Sep}
}

@article{szegedy2016inception,
  title={Inception-v4, inception-resnet and the impact of residual connections on learning},
  author={Szegedy, Christian and Ioffe, Sergey and Vanhoucke, Vincent and Alemi, Alex},
  journal={arXiv preprint arXiv:1602.07261},
  year={2016}
}

@article{nadipineni2020method,
  title={Method to Classify Skin Lesions using Dermoscopic images},
  author={Nadipineni, Hemanth},
  journal={arXiv preprint arXiv:2008.09418},
  year={2020}
}

@article{zunair2020melanoma,
  title={Melanoma detection using adversarial training and deep transfer learning},
  author={Zunair, Hasib and Hamza, A Ben},
  journal={Physics in Medicine \& Biology},
  year={2020},
  publisher={IOP Publishing}
}

@article{krizhevsky2017imagenet,
  title={Imagenet classification with deep convolutional neural networks},
  author={Krizhevsky, Alex and Sutskever, Ilya and Hinton, Geoffrey E},
  journal={Communications of the ACM},
  volume={60},
  number={6},
  pages={84--90},
  year={2017},
  publisher={ACM New York, NY, USA}
}

@article{esteva2017dermatologist,
  title={Dermatologist-level classification of skin cancer with deep neural networks},
  author={Esteva, Andre and Kuprel, Brett and Novoa, Roberto A and Ko, Justin and Swetter, Susan M and Blau, Helen M and Thrun, Sebastian},
  journal={nature},
  volume={542},
  number={7639},
  pages={115--118},
  year={2017},
  publisher={Nature Publishing Group}
}

@article{fornaciali2016towards,
  title={Towards automated melanoma screening: Proper computer vision \& reliable results},
  author={Fornaciali, Michel and Carvalho, Micael and Bittencourt, Fl{\'a}via Vasques and Avila, Sandra and Valle, Eduardo},
  journal={arXiv preprint arXiv:1604.04024},
  year={2016}
}

@article{masood2013computer,
  title={Computer aided diagnostic support system for skin cancer: a review of techniques and algorithms},
  author={Masood, Ammara and Ali Al-Jumaily, Adel},
  journal={International journal of biomedical imaging},
  volume={2013},
  year={2013},
  publisher={Hindawi}
}

@article{valle2020data,
  title={Data, depth, and design: Learning reliable models for skin lesion analysis},
  author={Valle, Eduardo and Fornaciali, Michel and Menegola, Afonso and Tavares, Julia and Bittencourt, Fl{\'a}via Vasques and Li, Lin Tzy and Avila, Sandra},
  journal={Neurocomputing},
  volume={383},
  pages={303--313},
  year={2020},
  publisher={Elsevier}
}

@incollection{perez2018data,
  title={Data augmentation for skin lesion analysis},
  author={Perez, F{\'a}bio and Vasconcelos, Cristina and Avila, Sandra and Valle, Eduardo},
  booktitle={OR 2.0 Context-Aware Operating Theaters, Computer Assisted Robotic Endoscopy, Clinical Image-Based Procedures, and Skin Image Analysis},
  pages={303--311},
  year={2018},
  publisher={Springer}
}

@article{yu2016automated,
  title={Automated melanoma recognition in dermoscopy images via very deep residual networks},
  author={Yu, Lequan and Chen, Hao and Dou, Qi and Qin, Jing and Heng, Pheng-Ann},
  journal={IEEE transactions on medical imaging},
  volume={36},
  number={4},
  pages={994--1004},
  year={2016},
  publisher={IEEE}
}

@incollection{bissoto2018skin,
  title={Skin lesion synthesis with generative adversarial networks},
  author={Bissoto, Alceu and Perez, F{\'a}bio and Valle, Eduardo and Avila, Sandra},
  booktitle={OR 2.0 Context-Aware Operating Theaters, Computer Assisted Robotic Endoscopy, Clinical Image-Based Procedures, and Skin Image Analysis},
  pages={294--302},
  year={2018},
  publisher={Springer}
}

@inproceedings{he2016deep,
  title={Deep residual learning for image recognition},
  author={He, Kaiming and Zhang, Xiangyu and Ren, Shaoqing and Sun, Jian},
  booktitle={Proceedings of the IEEE conference on computer vision and pattern recognition},
  pages={770--778},
  year={2016}
}

@article{simonyan2014very,
  title={Very deep convolutional networks for large-scale image recognition},
  author={Simonyan, Karen and Zisserman, Andrew},
  journal={arXiv preprint arXiv:1409.1556},
  year={2014}
}

@inproceedings{xu2015show,
  title={Show, attend and tell: Neural image caption generation with visual attention},
  author={Xu, Kelvin and Ba, Jimmy and Kiros, Ryan and Cho, Kyunghyun and Courville, Aaron and Salakhudinov, Ruslan and Zemel, Rich and Bengio, Yoshua},
  booktitle={International conference on machine learning},
  pages={2048--2057},
  year={2015}
}

@article{gessert2020skin,
  title={Skin lesion classification using ensembles of multi-resolution EfficientNets with meta data},
  author={Gessert, Nils and Nielsen, Maximilian and Shaikh, Mohsin and Werner, Ren{\'e} and Schlaefer, Alexander},
  journal={MethodsX},
  pages={100864},
  year={2020},
  publisher={Elsevier}
}

@article{rezvantalab2018dermatologist,
  title={Dermatologist level dermoscopy skin cancer classification using different deep learning convolutional neural networks algorithms},
  author={Rezvantalab, Amirreza and Safigholi, Habib and Karimijeshni, Somayeh},
  journal={arXiv preprint arXiv:1810.10348},
  year={2018}
}

@inproceedings{tran2015learning,
  title={Learning spatiotemporal features with 3d convolutional networks},
  author={Tran, Du and Bourdev, Lubomir and Fergus, Rob and Torresani, Lorenzo and Paluri, Manohar},
  booktitle={Proceedings of the IEEE international conference on computer vision},
  pages={4489--4497},
  year={2015}
}

@article{tomita2019attention,
  title={Attention-based deep neural networks for detection of cancerous and precancerous esophagus tissue on histopathological slides},
  author={Tomita, Naofumi and Abdollahi, Behnaz and Wei, Jason and Ren, Bing and Suriawinata, Arief and Hassanpour, Saeed},
  journal={JAMA network open},
  volume={2},
  number={11},
  pages={e1914645--e1914645},
  year={2019},
  publisher={American Medical Association}
}

@article{kingma2014adam,
  title={Adam: A method for stochastic optimization},
  author={Kingma, Diederik P and Ba, Jimmy},
  journal={arXiv preprint arXiv:1412.6980},
  year={2014}
}

@article{ioffe2015batch,
  title={Batch normalization: Accelerating deep network training by reducing internal covariate shift},
  author={Ioffe, Sergey and Szegedy, Christian},
  journal={arXiv preprint arXiv:1502.03167},
  year={2015}
}

@article{srivastava2014dropout,
  title={Dropout: a simple way to prevent neural networks from overfitting},
  author={Srivastava, Nitish and Hinton, Geoffrey and Krizhevsky, Alex and Sutskever, Ilya and Salakhutdinov, Ruslan},
  journal={The journal of machine learning research},
  volume={15},
  number={1},
  pages={1929--1958},
  year={2014},
  publisher={JMLR. org}
}

@inproceedings{huang2017densely,
  title={Densely connected convolutional networks},
  author={Huang, Gao and Liu, Zhuang and Van Der Maaten, Laurens and Weinberger, Kilian Q},
  booktitle={Proceedings of the IEEE conference on computer vision and pattern recognition},
  pages={4700--4708},
  year={2017}
}

@article{zhang2019attention,
  title={Attention residual learning for skin lesion classification},
  author={Zhang, Jianpeng and Xie, Yutong and Xia, Yong and Shen, Chunhua},
  journal={IEEE transactions on medical imaging},
  volume={38},
  number={9},
  pages={2092--2103},
  year={2019},
  publisher={IEEE}
}

@misc{wang2017residual,
      title={Residual Attention Network for Image Classification}, 
      author={Fei Wang and Mengqing Jiang and Chen Qian and Shuo Yang and Cheng Li and Honggang Zhang and Xiaogang Wang and Xiaoou Tang},
      year={2017},
      eprint={1704.06904},
      archivePrefix={arXiv},
      primaryClass={cs.CV}
}

@inproceedings{hu2018squeeze,
  title={Squeeze-and-excitation networks},
  author={Hu, Jie and Shen, Li and Sun, Gang},
  booktitle={Proceedings of the IEEE conference on computer vision and pattern recognition},
  pages={7132--7141},
  year={2018}
}

@inproceedings{Grad-cam,
  title={Grad-cam: Visual explanations from deep networks via gradient-based localization},
  author={Selvaraju, Ramprasaath R and Cogswell, Michael and Das, Abhishek and Vedantam, Ramakrishna and Parikh, Devi and Batra, Dhruv},
  booktitle={Proceedings of the IEEE international conference on computer vision},
  pages={618--626},
  year={2017}
}

@article{AUC,
  title={Using AUC and accuracy in evaluating learning algorithms},
  author={Huang, Jin and Ling, Charles X},
  journal={IEEE Transactions on knowledge and Data Engineering},
  volume={17},
  number={3},
  pages={299--310},
  year={2005},
  publisher={IEEE}
}

@article{yao2021single,
  title={Single Model Deep Learning on Imbalanced Small Datasets for Skin Lesion Classification},
  author={Yao, Peng and Shen, Shuwei and Xu, Mengjuan and Liu, Peng and Zhang, Fan and Xing, Jinyu and Shao, Pengfei and Kaffenberger, Benjamin and Xu, Ronald X},
  journal={arXiv preprint arXiv:2102.01284},
  year={2021}
}

@article{shen2021low,
  title={Low-cost and high-performance data augmentation for deep-learning-based skin lesion classification},
  author={Shen, Shuwei and Xu, Mengjuan and Zhang, Fan and Shao, Pengfei and Liu, Honghong and Xu, Liang and Zhang, Chi and Liu, Peng and Zhang, Zhihong and Yao, Peng and others},
  journal={arXiv preprint arXiv:2101.02353},
  year={2021}
}

@article{ham10000,
  title={The HAM10000 dataset, a large collection of multi-source dermatoscopic images of common pigmented skin lesions},
  author={Tschandl, Philipp and Rosendahl, Cliff and Kittler, Harald},
  journal={Scientific data},
  volume={5},
  number={1},
  pages={1--9},
  year={2018},
  publisher={Nature Publishing Group}
}

@article{ISIC-2017,
  author    = {Noel C. F. Codella and
               David Gutman and
               M. Emre Celebi and
               Brian Helba and
               Michael A. Marchetti and
               Stephen W. Dusza and
               Aadi Kalloo and
               Konstantinos Liopyris and
               Nabin K. Mishra and
               Harald Kittler and
               Allan Halpern},
  title     = {Skin Lesion Analysis Toward Melanoma Detection: {A} Challenge at the
               2017 International Symposium on Biomedical Imaging (ISBI), Hosted
               by the International Skin Imaging Collaboration {(ISIC)}},
  journal   = {CoRR},
  volume    = {abs/1710.05006},
  year      = {2017},
  url       = {http://arxiv.org/abs/1710.05006},
  archivePrefix = {arXiv},
  eprint    = {1710.05006},
  bibsource = {dblp computer science bibliography, https://dblp.org}
}
}
\end{document}